\documentclass[aps,10pt,pra,twocolumn,showpacs,nofootinbib,superscriptaddress]{revtex4-1}
\usepackage{amsfonts}
\usepackage{amsmath}
\usepackage{amssymb}
\usepackage{graphicx}%
\usepackage{color}
\usepackage{cancel}
\def\red{} 
\providecommand{\U}[1]{\protect\rule{.1in}{.1in}}
\providecommand{\U}[1]{\protect\rule{.1in}{.1in}}

\begin{document}
\title{Polarons in a Dipolar Condensate}
\author{Ben Kain}
\affiliation{Department of Physics, College of the Holy Cross, Worcester, Massachussets 01610, USA}
\affiliation{Department of Physics and Astronomy, Rowan University, Glassboro, New
Jersey 08028, USA}
\author{Hong Y.\ Ling}
\affiliation{Department of Physics and Astronomy, Rowan University, Glassboro, New
Jersey 08028, USA}
\affiliation{Kavli Institute for Theoretical Physics, University of California, Santa
Barbara, California 93106, USA }

\begin{abstract}
We consider a polaronic model in which impurity fermions interact with
background bosons in a dipolar condensate. The polaron in this model emerges
as an impurity dressed with a cloud of phonons of the dipolar condensate,
which, due to the competition between the attractive and repulsive part of the
dipole-dipole interaction, obey an anisotropic dispersion spectrum. We study
how this anisotropy affects the \v{C}erenkov-like emission of Bogoliubov
phonon modes, which can be directly verified by experiments in which a dipolar
BEC moves against an obstacle. We also study the spectral function of impurity
fermions, which is directly accessible to the momentum resolved rf
spectroscopy in cold atoms.

\end{abstract}

\pacs{67.85.-d, 67.85.Pq, 71.38.Fp}
\maketitle

\section{Introduction}

A conduction electron in an ionic crystal or a polar semiconductor displaces
nearby ions, thereby polarizing the medium in the vicinity of the
electron. An analogous picture emerges when an impurity atom is immersed in
an ultracold atomic quantum gas containing atoms distinct from but capable of
interacting with the impurity atom. The impurity finds itself surrounded by
and traveling with its local disturbance, a cloud of background atoms, forming
a polaron. In recent years, much effort has been focused on systems where
both impurity and background atoms are fermions (see \cite{chevy, massignan} for a review), inspired by the remarkable
agreement between theoretical predictions
\cite{prokofev08PhysRevB.77.020408,mora09PhysRevA.80.033607} and experimental
findings obtained from set-ups where a single spin-$\downarrow$ impurity is
immersed in a sea of spin-$\uparrow$ background atoms
\cite{schirotzek09PhysRevLett.102.230402,nascimbene09PhysRevLett.103.170402},
a setting that shares much resemblance to the Kondo problem in condensed
matter physics
\cite{anderson05CareerInTheoreticalPhysicsBook,hewson97HeavyFermionsBook}.
\ The present work, however, concerns polaron models where the impurity
atoms are fermions but the background atoms are bosons in a Bose-Einstein
condensate (BEC), in which density fluctuations of the BEC are described by
phonons. Recent years have also witnessed an increased interest in such
systems due largely to their similarity to the electron-phonon system where the
polaron picture \cite{landau33ZSoujetunion.3.644} is central to the
understanding of colossal magnetoresistance materials
\cite{mannella05Nature.428.474} and is believed to play a vital role in the
physics of high-$T_{c}$ superconductivity in strongly correlated materials
\cite{lanzara01Nature.412.6846,lee06RevModPhys.78.17,devreese09RepProgPhys.72.066501}
and in unconventional pairing mechanisms \cite{alexandrov08PhysRevB.77.094502}%
. 

In this paper, motivated by recent experimental advancement in achieving
dipolar quantum gases consisting of either heteronuclear molecules with
electric dipoles \cite{ni08Science.322.5899,ospelkaus09FaradayDiscuss142.351}
or atoms with magnetic dipoles
\cite{vengalattore08PhysRevLett.100.170403,stuhler05PhysRevLett.95.150406,lu11PhysRevLett.107.190401,aikawa12PhysRevLett.108.210401}%
, we consider the aforementioned Bose polaron models except that the
background atoms are now bosons in a dipolar condensate. In a dipolar quantum
gas \cite{santos00PhysRevLett.85.1791,yi00PhysRevA.61.041604}, the
dipole-dipole interaction represents a control knob inaccessible to nondipolar
bosons. Thus, mixing dipolar bosons with fermions opens up new possibilities.
An important consequence of the dipole-dipole interaction is that the phonon
spectrum of a dipolar condensate is no longer isotropic---one can tune the
dipolar interaction to lower the energy of a phonon along some directions
while simultaneously increasing it along other directions
\cite{lahaye09RepProgPhys.72.126401}, a phenomenon that a recent experiment
demonstrated using Raman-Bragg spectroscopy in a dipolar chromium BEC
\cite{bismut12PhysRevLett.109.155302}.\ Thus, impurity fermions submerged in a
dipolar condensate act as anisotropic polarons, interacting with surrounding
phonons. While many studies exist in the literature concerning the polaronic
Bose-Fermi models, including those in Refs.\ \cite{cucchietti06PhysRevLett.96.210401,tempere09PhysRevB.80.184504,casteels11PhysRevA.83.033631}
for large (continuous) polarons, those in Refs.\ \cite{bruderer07PhysRevA.76.011605,privitera10PhysRevA.82.063614} for small
(Holstein) polarons, and some experimental developments
\cite{chikkatur00PhysRevLett.85.483,catani12PhysRevA.85.023623,scelle13PhysRevLett.111.070401}%
, to the best of our knowledge, none of them have seriously considered
polaron systems with background atoms that are \textit{dipolar} bosons. The
purpose of our work here is to extend studies from nondipolar to dipolar
Bose-Fermi polaron models and develop theoretical tools which allow us to
gain, from investigations in the weak coupling limit, quantitative
understanding of the impurity polarons in the dipolar condensate.

In Sec.\ II, we present the Hamiltonian for the effective polaronic model where
density fluctuations of the dipolar condensate are described by phonons, and
we derive, in the same section, the impurity self-energy, taking into
consideration only single-phonon-impurity scattering processes. In Sec.\ III,
we evaluate this self-energy on the mass shell and use it to gain some
quantitative insight into the physics of the anisotropic polarons under
consideration. A moving polaron may emit phonons in much the same way that a
moving charge emits electromagnetic radiation or \v{C}erenkov radiation when
its velocity exceeds a certain threshold
\cite{cerenkov.2.451,cerenkov37PhysRev.52.378}. In addition to the effective
mass, the self-energy will be used in Sec.\ III to analyze the decay rate of
the polaron and to understand \v{C}erenkov-like phonon emissions from an
impurity in a dipolar condensate
\cite{muruganandam12PhysicsLettersA376.2012480}. In Sec.\ IV, we describe
polarons using Fermi liquid theory, focusing, however, on spectral functions,
which can, in principle, be probed using the momentum resolved rf spectroscopy
in cold atoms \cite{stewart08Nature.454.744,feld11Nature.480.75}. Finally,
we conclude in Sec.\ V.

\section{The Effective Hamiltonian and Impurity Self-Energy}

Let us now turn to the specific model of a cold-atom mixture in which
spin-polarized impurity fermions with mass $m_{F}$ permeate in a dipolar
condensate of bosons with mass $m_{B}$.  \red{The condensate is confined in a sufficiently large trap so that it is practically homogeneous.  In addition, all dipoles are assumed to point in the same direction as an external (either electric or magnetic) field, which we take to be the $z$ direction.}  The interactions in this model are
divided into a short-range part and a long-range part. The former is
described by $U_{BF}=4\pi\hbar^{2}a_{BF}/m_{BF}$ [$m_{BF}=2m_{B}m_{F}/\left(
m_{B}+m_{F}\right)  $] and $U_{BB}$ $=4\pi\hbar^{2}a_{BB}/m_{B}$, where
$a_{BF}$ and $a_{BB}$ are, respectively, impurity-boson and boson-boson s-wave
scattering lengths. The latter is described by $U_{DD}\left(  \mathbf{q}%
\right)  =8\pi d^{2}P_{2}\left(  z_{\mathbf{q}}\right)  $, which is the
dipole-dipole interaction between two bosons in momentum space, \red{with $d$ the
induced dipole moment, $P_{2}\left(  z_{\mathbf{q}}\right)  \equiv (3 z_\mathbf{q}^2 - 1)/2$ the second-order
Legendre polynomial, where $z_{\mathbf{q}} = \cos\theta_\mathbf{q}$ and $\theta_\mathbf{q}$ is the angle between
momentum $\hbar\mathbf{q}$ and the dipole direction (along the $z$ axis).} To
proceed, we limit our study to near zero temperature ($T\approx0$) where one
can approximate, within the Bogoliubov approximation, the dipolar Bose gas as
a uniform dipolar condensate of number density $n_{B}$ plus a collection of
phonons (due to density fluctuations) that obey the dispersion spectrum
\cite{goral00PhysRevA.61.051601,giovanazzi02PhysRevLett.89.130401}
\begin{equation}
\hbar\omega_{\mathbf{q}}=\hbar v_{B}q\sqrt{1+\left(  \xi_{B}q\right)
^{2}+2\varepsilon_{dd}P_{2}\left(  z_{\mathbf{q}}\right)  }, \label{E_K}%
\end{equation}
where $v_{B}=\sqrt{n_{B}U_{BB}/m_{B}}$ is the phonon speed in the absence of
the dipolar interaction, $\xi_{B}=\hbar/\sqrt{4m_{B}n_{B}U_{BB}}$ is the
healing length, and $\varepsilon_{dd}=4\pi d^{2}/(3U_{BB})$ measures the
dipolar interaction relative to the contact interaction $U_{BB}$. As such,
we are led to an effective Hamiltonian
\begin{align}
\hat{H}  &  =\sum_{\mathbf{p}}\left(  \xi_{\mathbf{p}}+n_{B}U_{BF}\right)
\hat{a}_{\mathbf{p}}^{\dag}\hat{a}_{\mathbf{p}}\nonumber\\
&  +\hbar\sum_{\mathbf{q}\neq0}\omega_{\mathbf{q}}\hat{\beta}_{\mathbf{q}%
}^{\dag}\hat{\beta}_{\mathbf{q}}+\frac{1}{\sqrt{V}}\sum_{\mathbf{q}\neq
0}g_{\mathbf{q}}\hat{\rho}_{\mathbf{q}}\left(  \hat{\beta}_{\mathbf{q}}%
+\hat{\beta}_{-\mathbf{q}}^{\dag}\right)  , \label{H}%
\end{align}
which describes impurity fermions (of field operator $\hat{a}_{\mathbf{p}}$
and density operator $\hat{\rho}_{\mathbf{q}}=\sum_{\mathbf{k}}\hat
{a}_{\mathbf{q}+\mathbf{k}}^{\dag}\hat{a}_{\mathbf{k}}$) interacting with
phonons (of field operator $\hat{\beta}_{\mathbf{q}}$) in a volume $V$. In
Eq.\ (\ref{H}), $\xi_{\mathbf{p}}=\epsilon_{\mathbf{p}}-\mu$ is the kinetic
energy of a fermion, $\epsilon_{\mathbf{p}}=\hbar^{2}p^{2}/2m_{F}$, relative
to the Fermi chemical potential $\mu$, and $g_{\mathbf{q}}=U_{BF}\sqrt
{n_{B}\hbar q^{2}/\left(  2m_{B}\omega_{\mathbf{q}}\right)  }$ is a
momentum-dependent coefficient measuring the impurity-phonon coupling
strength. In solid state systems, phonon dispersion due to lattice vibration
can be divided into an optical branch whose frequency remains almost
independent of the wavenumber and an acoustic branch whose frequency is
linearly proportional to the wavenumber \cite{kittel86SolidStatePhysics}.
Equation (\ref{H}) is the cold-atom analog of the Fr\"{o}hlich Hamiltonian
for electron-acoustic phonon systems since the Bogoliubov dispersion in Eq.\ (\ref{E_K}) asymptotes to that of an acoustic phonon
\cite{peeters85PhysRevB.32.3515} in the limit of long wavelength. In the
absence of the dipole-dipole interaction, Eq.\ (\ref{H}) has been employed,
within the context of cold atomic physics, to study polarons in the weak
coupling limit
\cite{casteels11PhysRevA.83.033631,casteels11PhysRevA.84.063612},
under\ self-localization
\cite{cucchietti06PhysRevLett.96.210401,santamore11NewJOfPhys.13.103029,blinova13arXiv:1304.7704}%
, and in the strong-coupling limit (using the Feynman path
integral) \cite{tempere09PhysRevB.80.184504,casteels12PhysRevA.86.043614}.
Having described our model in some detail, we now adopt a unit convention in
which $\hbar=1$ is implied (unless keeping $\hbar$ helps to elucidate the
physics).
\begin{figure}
[ptb]
\begin{center}
\includegraphics[
width=3.1in
]%
{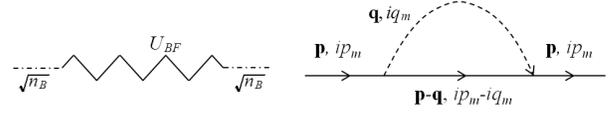}%
\caption{Feynman diagrams for the self energy of an impurity (fermion)
interacing with background bosons. }%
\label{Fig1:FeynmanDiagram}%
\end{center}
\end{figure}
\ 

\red{The polaron effect arises from the impurity-polaron interaction which modifies the self-energy and hence the bare propagator of the impurity.  Thus, much of the information about the properties of an impurity can be learned from its self-energy.  At zero temperature, this self-energy can be conveniently obtained from the nonzero-temperature Matsubara results by setting $T=0$.}  In the weak coupling regime, the Matsubara self-energy for impurity fermions,
up to second order in the Bose-Fermi interaction $U_{BF}$, reads
\begin{align}
\Sigma\left(  \mathbf{p},i\omega_{n}\right)   &  =U_{BF}n_{B}-\frac{1}{\beta
V}\sum_{\mathbf{q},iq_{m}}g_{\mathbf{q}}^{2}\times\nonumber\\
&  D_{0}\left(  \mathbf{q},iq_{m}\right)  G_{0}\left(  \mathbf{p}%
-\mathbf{q},i\omega_{n}-iq_{m}\right)  , \label{self-energy 0}%
\end{align}
where the first term is due to the interaction of the impurity fermions with
the condensed bosons [Fig.\ref{Fig1:FeynmanDiagram}(left)], and the second
term accounts for all single-phonon-impurity scattering processes and can thus
be read off directly from the Feynman diagram involving one phonon [Fig. \ref{Fig1:FeynmanDiagram}(right)]. In Eq.\ (\ref{self-energy 0}),
$G_{0}\left(  \mathbf{p},i\omega_{n}\right)  =\left(  i\omega_{n}%
-\xi_{\mathbf{p}}\right)  ^{-1}$ and $D_{0}\left(  \mathbf{q},iq_{m}\right)
=2\omega_{\mathbf{q}}[  (  iq_{m})  ^{2}-\omega_{\mathbf{q}%
}]  ^{-1}$ are the Matsubara Green's functions for noninteracting
fermions and phonons, respectively, with $\omega_{n}$ and $q_{m}$ being
fermion and boson Matsubara frequencies, which are odd and even multiples of
$\pi/\beta=\pi k_{B}T$, respectively. Analytic continuation of
$\Sigma\left(  \mathbf{p},i\omega_{n}\right)  $ leads to the retarded
self-energy $\Sigma^{R}\left(  \mathbf{p},\omega\right)  =\Sigma\left(
\mathbf{p},ip_{n}\rightarrow\omega+i0^{+}\right)  $, which, at zero
temperature and in the single (attractive) impurity limit, is given by
\begin{align}
\Sigma^{R}\left(  \mathbf{p},\omega\right)   &  =U_{BF}n_{B}+\frac{1}{V}%
\sum_{\mathbf{q}}\times\nonumber\\
&  \left(  \frac{g_{\mathbf{q}}^{2}}{\omega-\xi_{\mathbf{p}-\mathbf{q}}%
-\omega_{\mathbf{q}}+i0^{+}}+\frac{m_{FB}U_{FB}^{2}}{2n_{B}^{-1}m_{F}%
\epsilon_{\mathbf{q}}}\right)  . \label{self-energy}%
\end{align}
In leading to Eq.\ (\ref{self-energy}), we have omitted an important step which
we now explain. It goes back to Eqs.\ (\ref{H}) and (\ref{self-energy 0})
where $U_{BF}$\ should really be written as $U_{BF}^{\prime}$, a parameter
introduced to model the Bose-Fermi interaction; our use of $U_{BF}$ instead of
$U_{BF}^{\prime}$ is to simplify notation. $\ U_{BF}^{\prime}$ is\ connected
to $U_{BF}$, up to the second order in$\ U_{BF}$, through
\cite{tempere09PhysRevB.80.184504} \
\begin{equation}
U_{BF}^{\prime}=U_{BF}+\frac{U_{BF}^{2}}{V}\sum_{\mathbf{q}\neq0}\frac{m_{BF}%
}{q^{2}}, \label{Lippmann}%
\end{equation}
according to the Lippmann-Schwinger equation for the two-body vacuum T-matrix
\cite{fetter71ManyParticleSystemsBook} that describes the scattering between a
boson and a fermion. Upon its substitution in Eq.\ (\ref{self-energy 0}), the
linear term in Eq.\ (\ref{Lippmann}) becomes the first term in Eq.
(\ref{self-energy}) while the second order term in Eq.\ (\ref{Lippmann}) is the
origin of the last term in Eq.\ (\ref{self-energy}), which serves as a
renormalization factor regulating the ultraviolet divergence known to exist in
the integration for the self energy in Eq.\ (\ref{self-energy 0}).

This self-energy [Eq.\ (\ref{self-energy})], which, strictly speaking, is valid
only in the weak coupling regime, allows us to explore the subject in a
conceptually intuitive and mathematically simplified manner. The ensuing
results are expected to capture all qualitative features about the anisotropic
polarons, provided that the impurity-phonon coupling is still much lower than
the critical value for self-localization
\cite{cucchietti06PhysRevLett.96.210401,santamore11NewJOfPhys.13.103029,blinova13arXiv:1304.7704,kalas06PhysRevA.73.043608,bruderer08EurPhysLett.82.30004}%
.

In what follows, we will use Eq.\ (\ref{self-energy}) or equivalently%
\begin{align}
\operatorname{Re}\Sigma^{R}\left(  \mathbf{p},\omega\right)   &  =U_{BF}%
n_{B}+\int\frac{d^{3}\mathbf{q}}{\left(  2\pi\right)  ^{3}}\times\nonumber\\
&  \left(  \mathcal{P}\frac{g_{\mathbf{q}}^{2}}{\omega-\xi_{\mathbf{p}%
-\mathbf{q}}-\omega_{\mathbf{q}}}+\frac{m_{FB}U_{FB}^{2}}{2n_{B}^{-1}%
m_{F}\epsilon_{\mathbf{q}}}\right)  , \label{self-energy real}%
\end{align}
and
\begin{equation}
\operatorname{Im}\Sigma^{R}\left(  \mathbf{p},\omega\right)  =-\pi\int
\frac{d^{3}\mathbf{q}}{\left(  2\pi\right)  ^{3}}g_{\mathbf{q}}^{2}%
\delta\left(  \omega-\xi_{\mathbf{p}-\mathbf{q}}-\omega_{\mathbf{q}}\right)  ,
\label{self-energy imaginary}%
\end{equation}
to study polarons in the dipolar condensate where $\mathcal{P}$ stands for the
principal value.
\begin{figure}
[ptb]
\begin{center}
\includegraphics[
width=3.3in
]%
{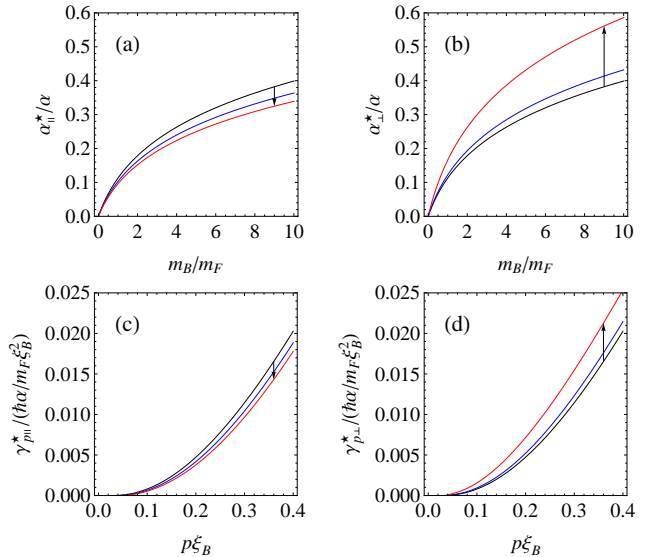}%
\caption{\red{(Color online) (a) $\alpha_{||}$ and (b) $\alpha_{\perp}$
as a function of $m_{B}/m_{F}$ for different $\varepsilon_{dd}=0$ (black), 0.3
(blue), and 0.8 (red). (c) $\gamma_{p_{\parallel}}^{\star}$ and (d)
$\gamma_{p_{\perp}}^{\star}$ as a function of $p$ for different
$\varepsilon_{dd}$ when $m_{B}/m_{F}$ is fixed to $21$.  Note that because of the rich existence of atomic elements and their isotopes in nature, for practical purposes one can regard the mass ratio as a tunable parameter.  The special case of $m_{B}/m_{F} = 21$ in (c) and (d) corresponds to the mass ratio in the $^{6}$Li and $^{40}$K$^{86}$Rb mixture.}}
\label{Fig2:On-Mass-ShellResults}%
\end{center}
\end{figure}

\section{Polaron Description on the Mass Shell and \v{C}erenkov type of phonon
emissions}

In order to build some analytical insight, we first estimate the self-energy
using the polaron energy $\omega$ on the mass shell, $\omega=\xi_{\mathbf{p}}%
$, and use this self-energy to characterize the polarons of interest. In
this limit, we find that at small momenta where $\operatorname{Im}\Sigma
^{R}\left(  \mathbf{p},\xi_{\mathbf{p}}\right)  =0$, the impurity energy%
\begin{equation}
E_{\mathbf{p}}=\epsilon_{\mathbf{p}}+\operatorname{Re}\Sigma^{R}\left(
\mathbf{p},\xi_{\mathbf{p}}\right)  , \label{Ep}%
\end{equation}
is approximated as%
\begin{equation}
E_{\mathbf{p}}\approx E_{0}^{\star}+p_{\parallel}^{2}/2m_{F||}^{\star
}+p_{\perp}^{2}/2m_{F\perp}^{\star}, \label{Ep approx}%
\end{equation}
where $p_{\parallel}$ and $p_{\perp}$ are, respectively, the momentum
component along the dipole direction (axial momentum) and that normal to the
dipole direction (radial momentum), and $m_{F||}^{\star}$ and $m_{F\perp
}^{\star}$ are, respectively, the axial and radial effective polaron mass, the
only two surviving elements in the mass tensor. Let $\bar{q}=q\xi_{B}$,
$\bar{p}=p\xi_{B}$, $\bar{m}_{B}=m_{B}/m_{F}$, and $\bar{\omega}_{\mathbf{q}%
}=\omega_{\mathbf{q}}/\left(  \hbar/m_{F}\xi_{B}^{2}\right)  $ or
\begin{equation}
\bar{\omega}_{\mathbf{q}}=\bar{q}\sqrt{1+\bar{q}^{2}+2\varepsilon_{dd}%
P_{2}\left(  z_{\mathbf{q}}\right)  }/\left(  2\bar{m}_{B}\right)  ,
\end{equation}
be the scaled quantities corresponding to $q$, $p$, $m_{B}$, and
$\omega_{\mathbf{q}}$ in the so-called polaron unit system. Taylor expanding
Eq.\ (\ref{self-energy real}) up to the second order in the momenta, we find
that
\begin{equation}
m_{F||}^{\star}=m_{F}/\left(  1-\alpha_{||}^{\star}\right), \quad
m_{F\perp}^{\star}=m_{F}/\left(  1-\alpha_{\perp}^{\star}\right)  ,
\end{equation}
where
\begin{subequations}
\label{alpha star}%
\begin{align}
\alpha_{\parallel}^{\star}  &  =2\alpha_{\Sigma}\int d\bar{q}\int_{-1}%
^{+1}dz_{\mathbf{q}}\frac{\bar{q}^{6}z_{\mathbf{q}}^{2}}{\bar{\omega
}_{\mathbf{q}}\left(  \bar{\omega}_{\mathbf{q}}+\bar{q}^{2}/2\right)  ^{3}},\\
\alpha_{\perp}^{\star}  &  =\alpha_{\Sigma}\int d\bar{q}\int_{-1}%
^{+1}dz_{\mathbf{q}}\frac{\bar{q}^{6}\left(  1-z_{\mathbf{q}}^{2}\right)
}{\bar{\omega}_{\mathbf{q}}\left(  \bar{\omega}_{\mathbf{q}}+\bar{q}%
^{2}/2\right)  ^{3}}.
\end{align}
In Eqs.\ (\ref{alpha star}),%
\end{subequations}
\begin{equation}
\alpha_{\Sigma}=\frac{\alpha}{32\pi}\frac{1}{\bar{m}_{B}}\left(  1+\frac
{1}{\bar{m}_{B}}\right)  ^{2},
\end{equation}
is the coefficient expressed in terms of
\begin{equation}
\alpha=a_{BF}^{2}/\left(  a_{BB}\xi_{B}\right)  , \label{alpha}%
\end{equation}
which is a (dimensionless) polaronic coupling constant
\cite{tempere09PhysRevB.80.184504,casteels11PhysRevA.84.063612}. The ability
to enhance $\alpha$ in Eq.\ (\ref{alpha}) by tuning s-wave scattering lengths
via Feshbach resonance
\cite{tiesinga93PhysRevA.47.4114,inouye98Nature.392.151,timmermans99PhysRep.315.199}
has been the main motivation behind the recent upsurge of activity in
exploring polaronic physics in Bose-Fermi mixtures of cold atoms.

\ Figure \ref{Fig2:On-Mass-ShellResults} (a) and (b) illustrate how
$\alpha_{\perp}$ and $\alpha_{||}$ change with the mass ratio, $m_{B}/m_{F}$,
for different\ values of the dipolar interaction $\varepsilon_{dd}$. As a
function of $m_{B}/m_{F}$ (for fixed $\varepsilon_{dd}$), $\alpha_{||}$
behaves similarly to $\alpha_{\perp}$---both increase appreciably with
$m_{B}/m_{F}$. As a function of $\varepsilon_{dd}$ (for fixed $m_{B}/m_{F}%
$), $\alpha_{||}$ behaves oppositely to $\alpha_{\perp}$---as $\varepsilon
_{dd}$ increases (but staying less than 1, beyond which the dipolar condensate is
unstable), $\alpha_{||}$ decreases while
$\alpha_{\perp}$ increases.  To account for this
anisotropy, we first note from Eq.\ (\ref{alpha star}) that contributions to
$\alpha_{||}$ and $\alpha_{\perp}$ stem primarily from axial phonons (with
$z_{\mathbf{q}}$ being close to 1) and radial phonons (with $z_{\mathbf{q}}$
being close to $0$), respectively. In a dipolar condensate in which the
attractive (head-to-tail) interaction competes with the repulsive
(side-to-side) interaction, tuning the dipolar interaction $\varepsilon_{dd}$
(towards 1)\ serves to reduce the energy cost of the radial phonons while
simultaneously increasing the energy cost of the axial phonons.  As a result, at
a very low impurity energy, such a tuning amounts to resonantly enhancing the
interaction between the impurity and the radial phonon, on one hand, and
off-resonantly suppressing the interaction between the impurity and the axial
phonon, on the other hand. In turn, this results in $\alpha_{||}$ and
$\alpha_{\perp}$ changing in the opposite directions in response to the change
in the dipolar interaction, leading to the mass anisotropy in the polarons.

Let us now turn our attention to the imaginary part of the self-energy,
$\operatorname{Im}\Sigma^{R}\left(  \mathbf{p},\xi_{\mathbf{p}}\right)  ,$ on
the mass shell and use it to obtain the spontaneous decay rate of an impurity
polaron (in the sense of Fermi's golden rule), $\gamma_{\mathbf{p}}^{\star}=-$
$\operatorname{Im}\Sigma^{R}\left(  \mathbf{p},\xi_{\mathbf{p}}\right)  $.
\ Physically, when the imaginary part of the self-energy is non-vanishing, the
energy of the impurity dissipates by spontaneous emission of phonons in the
same sense of \v{C}erenkov radiation
\cite{cerenkov.2.451,cerenkov37PhysRev.52.378}, where a charged particle emits
electromagnetic radiation, when moving at a velocity higher than the phase
velocity of light in a dispersive background medium. The speed of the phonon
thus created depends on $z_{\mathbf{q}}$, the cosine of the angle between
phonon momentum $\mathbf{q}$ and the dipole direction, according to
\begin{equation}
v_{B}\left(  z_{\mathbf{q}}\right)  =\lim_{\left\vert \mathbf{q}\right\vert
\rightarrow0}\frac{\omega_{\mathbf{q}}}{\left\vert \mathbf{q}\right\vert
}=v_{B}\sqrt{1+2\varepsilon_{dd}P_{2}\left(  z_{\mathbf{q}}\right)  }.
\label{sound velocity}%
\end{equation}
The momentum of the emitted phonon, dictated by energy conservation as
represented by the Dirac $\delta$-function in Eq.\ (\ref{self-energy imaginary}%
), has magnitude%
\begin{align}
\bar{q}  &  =\frac{2\bar{p}y_{\mathbf{q}}}{1-\bar{m}_{B}^{-2}}-\nonumber\\
&  \frac{\sqrt{\left(  2\bar{p}y_{\mathbf{q}}\right)  ^{2}+\left(  1-\bar
{m}_{B}^{-2}\right)  \left(  1+2\varepsilon_{dd}P_{2}\left(  z_{\mathbf{q}%
}\right)  \right)  }}{\bar{m}_{B}\left(  1-\bar{m}_{B}^{-2}\right)  },
\label{q^-}%
\end{align}
which is a function of not only $z_{\mathbf{q}}$ but also $y_{\mathbf{q}}$,
the cosine of the angle between phonon momentum $\mathbf{q}$ and impurity
momentum $\mathbf{p}$.
\begin{figure}
[ptb]
\begin{center}
\includegraphics[
width=3.4in
]%
{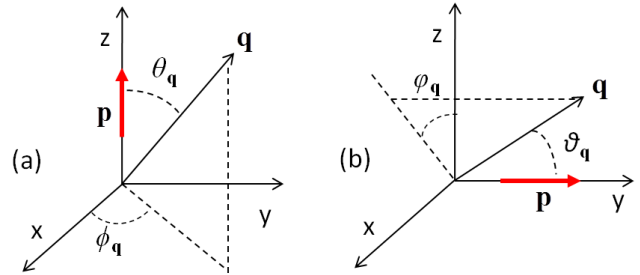}%
\caption{(Color online) (a) The spherical coordinate system in which the
zenith points to the $z$ axis and (b) the spherical coordinate system in which
the zenith points to the $y$ axis. (a) and (b) are used to study phonon
emissions by an impurity with momentum (thick red arrow) along the axial and
radial directions, respectively.}%
\label{Fig:SphericalCoordinates}%
\end{center}
\end{figure}
\ 

The anisotropy in the emitted phonons leads to the anisotropy in the decay
rates. For an axial polaron, we use the usual spherical coordinate\ system
in which the zenith points to the $z$ direction along which the axial impurity
moves [Fig.\ \ref{Fig:SphericalCoordinates}(a)]. $\ $\red{In this coordinate system, since both the momentum and dipoles are oriented along the $z$ direction,
$z_{\mathbf{q}}$ $=y_{\mathbf{q}}=\cos\theta_{\mathbf{q}}$, where $\left(
\theta_{\mathbf{q}},\phi_{\mathbf{q}}\right)  $ are the usual polar and
azimuthal angles.} An analysis of $\operatorname{Im}\Sigma^{R}\left(
\mathbf{p},\xi_{\mathbf{p}}\right)  $ then finds the axial decay rate to be
given by

\begin{align}
\gamma_{p_{\mathbf{\parallel}}}  &  =\frac{\hbar}{m_{F}\xi_{B}^{2}}\pi
\alpha_{\Sigma}\Theta\left(  \bar{p}-\frac{\sqrt{1+2\varepsilon_{dd}}}%
{2\bar{m}_{B}}\right)  \times\nonumber\\
&  \int_{\cos\theta_{m}}^{1}\frac{\bar{q}^{5}dz_{\mathbf{q}}}{\left\vert
\left(  \bar{q}-\bar{p}y_{\mathbf{q}}\right)  \bar{q}\omega_{\mathbf{q}%
}+\omega_{\mathbf{q}}^{2}+\frac{\bar{q}^{4}}{\left(  2\bar{m}_{B}\right)
^{2}}\right\vert }, \label{axial}%
\end{align}
where $\theta_{m}$ will be defined in Eq.\ (\ref{theda axial}). \red{For a radial}
polaron, we adopt a different spherical coordinate system where the zenith
points to the $y$ direction along which the radial impurity is assumed to move
[Fig.\ \ref{Fig:SphericalCoordinates}(b)]. In such a coordinate system,
$y_{\mathbf{q}}=\cos\vartheta_{\mathbf{q}}$ and $z_{\mathbf{q}}=\sin
\vartheta_{\mathbf{q}}\cos\varphi_{\mathbf{q}}$, where $\left(  \vartheta
_{\mathbf{q}}\text{,}\varphi_{\mathbf{q}}\right)  $ are the polar and
azimuthal angle of this new coordinate system. A similar analysis of
$\operatorname{Im}\Sigma^{R}\left(  \mathbf{p},\xi_{\mathbf{p}}\right)  $ then
finds the radial decay rate to be given by
\begin{align}
\gamma_{p_{\mathbf{\perp}}}  &  =\frac{\hbar}{m_{F}\xi_{B}^{2}}\pi
\alpha_{\Sigma}\Theta\left(  \bar{p}-\frac{\sqrt{1-\varepsilon_{dd}}}{2\bar
{m}_{B}}\right)  \int_{0}^{2\pi}\frac{d\varphi_{\mathbf{q}}}{2\pi}\nonumber\\
&  \times\int_{\cos\vartheta_{m}}^{1}\frac{\bar{q}^{5}dy_{\mathbf{q}}%
}{\left\vert \left(  \bar{q}-\bar{p}y_{\mathbf{q}}\right)  \bar{q}\bar{\omega
}_{\mathbf{q}}+\bar{\omega}_{\mathbf{q}}^{2}+\frac{\bar{q}^{4}}{\left(
2\bar{m}_{B}\right)  ^{2}}\right\vert }, \label{radial}%
\end{align}
where $\vartheta_{m}$ will be defined in Eq.\ (\ref{theda radial}).
Again,$\ \bar{q}$ in both Eqs.\ (\ref{axial}) and (\ref{radial}) is defined in
Eq.\ (\ref{q^-}), and has the physical meaning of being the magnitude of the
momentum of an emitted phonon. Figure \ref{Fig2:On-Mass-ShellResults} (c)
and (d) plot $\gamma_{p_{\parallel}}^{\star}$ and $\gamma_{p_{\bot}}^{\star}$
as functions of $p$. The physical meanings of other terms are explained as
follows. 

The Heaviside step function, $\Theta\left(  x\right)  $, implements the Laudau
criterion for superfluidity where the elementary excitations are the
anisotropic phonons of a dipolar condensate. Thus, for an impurity traveling
along the axial direction, dissipation takes place only when its velocity $v$
exceeds the sound velocity at $z_{\mathbf{q}}=1$ or $v_{B} ( 1)
=v_{B}\sqrt{1+2\varepsilon_{dd}}$ [Eq.\ (\ref{sound velocity})], a condition
equivalent to $\bar{p}>\sqrt{1+2\varepsilon_{dd}}/2\bar{m}_{B}$ in the polaron
unit system. A similar analysis shows that for an impurity traveling along
the radial direction where $z_{\mathbf{q}}=0$, dissipation takes place only
when $\bar{p}>\sqrt{1-\varepsilon_{dd}}/2\bar{m}_{B}$. 

The angles, $\theta_{m}$ in Eq.\ (\ref{theda axial}) and $\vartheta_{m}$ in Eq.
(\ref{theda radial}), have the physical meaning that the phonon modes emitted
by the supersonic impurity along the axial $(z)$ direction are within the cone
of the polar angle (relative to the $z$-axis)
\begin{equation}
\theta_{m}=\cos^{-1}\sqrt{\frac{1-\varepsilon_{dd}}{\left(  2\bar{p}\bar
{m}_{B}\right)  ^{2}-3\varepsilon_{dd}}}, \label{theda axial}%
\end{equation}
and those emitted by the supersonic impurity along the radial direction are
within the cone of the polar angle (relative to the $y$-axis)%

\begin{equation}
\vartheta_{m}=\cos^{-1}\sqrt{\frac{1-\varepsilon_{dd}+3\varepsilon_{dd}%
\sin^{2}\varphi_{\mathbf{q}}}{\left(  2\bar{p}\bar{m}_{B}\right)
^{2}+3\varepsilon_{dd}\sin^{2}\varphi_{\mathbf{q}}}}. \label{theda radial}%
\end{equation}
\ 

As one may verify, the sine of the angle $\pi/2-\theta_{m}$ ($\pi
/2-\vartheta_{m}$) equals the ratio of the speed of sound to that of an axial
(radial) polaron. Thus, by definition, $\pi/2-\theta_{m}$ ($\pi
/2-\vartheta_{m}$) is the Mach angle, or equivalently $\cos\theta_{m}$
($\cos\vartheta_{m}$) is the inverse of the Mach number
\cite{machSitzungsberichte.95.764},\ which is a figure of merit well-known in
the study of shock waves, examples of which include sonic booms generated by
supersonic flights \cite{machSitzungsberichte.95.764}, bow and stern waves
created by high speed boats \cite{whitham74Book}, \v{C}erenkov radiation
\cite{cerenkov.2.451,cerenkov37PhysRev.52.378} emitted by charged particles,
and, more recently, Bogoliubov phonons emitted by atoms in superfluids
\cite{damski04PhysRevA.69.043610,kulikov03PhysRevA.67.063605,kamchatnov04PhysRevA.69.063605,salasnich07PhysRevA.75.043616,carusotto06PhysRevLett.97.260403,gladush07PhysRevA.75.033619}%
. In contrast to the axial impurity which preserves the cylindrical
symmetry so that the cone of the polar angle $\theta_{m}$ [Eq.
(\ref{theda axial})] is independent of the azimuthal angle, the radial
impurity breaks the cylindrical symmetry so that the cone of the polar angle
$\vartheta_{m}$ [Eq.\ (\ref{theda radial})] is now a function of azimuthal
angle $\varphi_{\mathbf{q}}$, which can only occur in systems with anisotropy.

\ The invariance under Galilean transformations implies that one can test
this anisotropy by letting a dipolar condensate move against a stationary
impurity (for example, in the form of a localized optical potential of a
far-detuned laser beam \cite{amo11Science332.03062011,cornell04}). The
emitted phonons are expected to form, behind the impurity, a conical wave
front of \textit{fixed} aperture $\pi/2-\theta_{m}$ when moving along the
axial direction and that of \textit{varying} aperture $\pi/2-\vartheta_{m}$
[in the fashion of Eq.\ (\ref{theda radial})] when moving along the radial
direction \cite{muruganandam12PhysicsLettersA376.2012480}.

\section{Spectral Functions and Polarons According to Fermi Liquid Theory}

We propose to probe the polarons in BECs using the same rf spectroscopy that
was successfully applied to explore the BEC-BCS crossover
\cite{chin04Science305.20082004} in resonant Fermi gases
\cite{kinnunen04Science305.20082004} and fermionic polaron physics in highly
imbalanced Fermi gases
\cite{schirotzek09PhysRevLett.102.230402,feld11Nature.480.75}. In rf
spectroscopy, an rf field of amplitude $\Omega_{L}$ and frequency $\omega_{L}$
is applied to promote impurities to the excited (final) hyperfine state
$| e \rangle $ that lies above the impurity state $|g\rangle $ by an energy $\hbar\omega_{eg}$, a process described by the
Hamiltonian
\begin{equation}
\hat{H}_{rf}=\frac{\hbar\Omega_{L}}{2}\int d^{3}\mathbf{r}\left[
e^{-i\omega_{L}t}\hat{\psi}_{e}^{\dag}\left(  \mathbf{r}\right)  \hat{\psi
}_{g}\left(  \mathbf{r}\right)  +h.c.\right]  ,
\end{equation}
where $\hat{\psi}_{g,e}\left(  \mathbf{r}\right)  $ are the relevant
annihilation operators. The rf signal, which is the rate at which the
population in the final state changes, is given by%
\begin{equation}
\ I=-2\left(  \frac{\Omega_{L}}{2}\right)  ^{2}\operatorname{Im}\chi\left(
\mu-\mu_{e}-\omega_{L}\right)  ,
\end{equation}
within linear response theory \red{(in which $\hat{H}_{rf}$ is treated as a small perturbation),} where $\mu_{e}$ is the chemical potential of
the final state and $\operatorname{Im}\chi(  \omega)  $ is the
Fourier transform of the retarded time-ordered correlation function,
$-i\Theta(  t-t^{\prime})  \langle [  \hat{\psi}%
_{e}^{\dag}(  \mathbf{r},t)  \psi_{g}(  \mathbf{r},t)
,\hat{\psi}_{g}^{\dag}(  \mathbf{r}^{\prime},t^{\prime})  \psi
_{e}(  \mathbf{r}^{\prime},t^{\prime})  ]  \rangle $,
averaged over both $\mathbf{r}$ and $\mathbf{r}^{\prime}$
\cite{punk07PhysRevLett.99.170404,massignan08PhysRevA.77.031601}. In the
case where the final state interaction is negligible, one can express, within
the Matsubara (imaginary-time) formalism, the rf signal as $I(
\omega)  \propto\int d^{3}\mathbf{p}\, I_{0}(  \mathbf{p}%
,\omega)  $ in terms of the momentum-resolved current
\begin{equation}
I_{0}\left(  \mathbf{p},\omega\right)  =\left(  \frac{\Omega_{L}}{2}\right)
^{2}A\left(  \mathbf{p},\xi_{\mathbf{p}}-\omega\right)  f\left(
\xi_{\mathbf{p}}-\omega\right)  ,
\end{equation}
where $\omega=\omega_{L}-\omega_{eg}$ is the rf detuning, $f\left(  x\right)
=(  e^{\beta x}+1)  ^{-1}$ is the Fermi distribution function which
becomes a step function $\Theta(  -x)  $ in the limit of zero
temperature, and $A\left(  \mathbf{p},\omega\right)  $ is the spectral
function satisfying the sum rule $\int_{-\infty}^{+\infty}d\omega \, A(
\mathbf{p},\omega)  /2\pi=1$. The momentum resolved rf spectroscopy
\cite{stewart08Nature.454.744,feld11Nature.480.75} is the cold-atom analog of
the angle resolved photoemission spectroscopy in solid state physics
\cite{damascelli03RevModPhys.75.473}. It measures $I_{0}\left(
\mathbf{p},\omega\right)  $ and thus allows the spectral function, $A\left(
\mathbf{p},\omega\right)  =-2\operatorname{Im}G\left(  \mathbf{p}%
,\omega+i0^{+}\right)  ,$ to be accessed experimentally, where $G\left(
\mathbf{p},i\omega_{n}\right)  $ is the Matsubara Green's function determined
by Dyson's equation $G^{-1}\left(  \mathbf{p},i\omega_{n}\right)
=G_{0}^{-1}\left(  \mathbf{p},i\omega_{n}\right)  -\Sigma\left(
\mathbf{p},i\omega_{n}\right)  .$ In terms of the retarded self-energy, the
spectral function $A\left(  \mathbf{p},\omega\right)  $ takes the form%
\begin{equation}
A\left(  \mathbf{p},\omega\right)  =\frac{-2\operatorname{Im}\Sigma^{R}\left(
\mathbf{p},\omega\right)  }{\left[  \omega-\xi_{\mathbf{p}}-\operatorname{Re}%
\Sigma^{R}\left(  \mathbf{p},\omega\right)  \right]  ^{2}+\operatorname{Im}%
\Sigma^{R}\left(  \mathbf{p},\omega\right)  ^{2}}.
\end{equation}
\begin{figure}
[ptb]
\begin{center}
\includegraphics[
width=3.3in
]%
{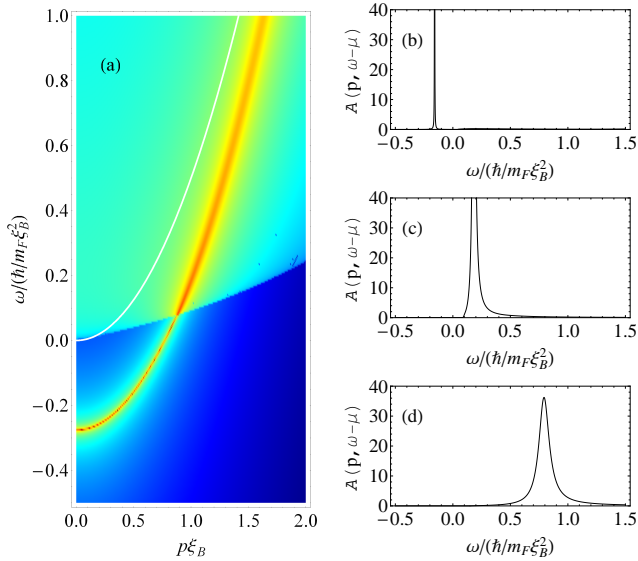}%
\caption{(Color online) (a) The contour plot of the spectral function [log($A\left(
\mathbf{p},\omega-\mu\right)  $)] of axial plarons$\ $in the $\left(
p,\omega\right)  $ space. The free Fermi dispersion relation is plotted as a
thin white line. The figures on the right panel show how the spectral function
change with $\omega$ for polarons with axial momentum of 0.5 (b), 1 (c), and
1.5 (d).\ The parameters used are $\varepsilon_{dd}=0.95,$ $m_{F}=6$ $u$,
$m_{B}=10m_{F}$, $a_{BB}=200a_{0}$, $a_{BF}=-500a_{0}$, $n_{B}=10^{20}$
m$^{-3}$, where $u$ is the atomic mass and $a_{0}$ is the Bohr radius. Note
that in this work, we are interested in the physics of mobile impurities.
\ Thus,\ $\alpha$ is chosen to be below the critical value for self
localization. In the considered example, $\alpha=0.48$. In producing these
diagrams, a small but finite number is used to approximate $0^{+}$.}%
\label{Fig3:spectralFunction}%
\end{center}
\end{figure}

Figure \ref{Fig3:spectralFunction}(a) showcases a typical contour plot of
$A\left(  \mathbf{p},\omega-\mu\right)  $ for an axial impurity. The key
features of this diagram can be captured in terms of the Fermi-liquid
parameters. The peak position, $E_{\mathbf{p}}$, is the root to the
equation,
\begin{equation}
E_{\mathbf{p}}-\mu=\xi_{\mathbf{p}}+\operatorname{Re}\Sigma^{R}\left(
\mathbf{p},E_{\mathbf{p}}-\mu\right)  ,
\end{equation}
involving the real part of the self-energy. In the region close to
$\mathbf{p}=0$, the peak position is approximated as a quadratic function of
momenta, $E_{\mathbf{p}}\approx E_{0}+p_{\parallel}^{2}/2m_{F||}^{\ast
}+p_{\perp}^{2}/2m_{F\perp}^{\ast}$. The curvature close to $\mathbf{p}=0$
is inversely proportional to the effective mass, $m_{F||\text{,}\perp}^{\ast
}=m_{F}Z_{0}^{-1}(  1+\partial_{\epsilon_{p_{\parallel}\text{,}_{\perp}}}\operatorname{Re}\Sigma^{R}\left(  \mathbf{p},\omega\right)  )  ^{-1}$,
where $Z_{0}=(  1-\partial_{\omega}\operatorname{Re}\Sigma^{R}\left(
\mathbf{p},\omega\right)  )  ^{-1}$ is the quasiparticle residue,
$\epsilon_{p_{\mathbf{\parallel}\text{,}\perp}\text{ }}=p_{\parallel
\text{,}\perp}^{2}/2m_{F\text{ }}$ is the kinetic energy, and every term in
$m_{F||\text{,}\perp}^{\ast}$ is evaluated at $\mathbf{p}=\omega=0$. 

The $\omega$-intercept of the peak line in Fig.\ \ref{Fig3:spectralFunction}(a)
with the vertical axis determines the ground state polaron energy $E_{0}$.
 In the single polaron limit, $E_{0}=\mu,$ where the chemical potential,
$\mu,$ is determined self consistently from $\mu=\operatorname{Re}%
\Sigma\left(  \mathbf{0},0\right)  $ \cite{combescot07PhysRevLett.98.180402}
or
\begin{equation}
\mu=U_{BF}n_{B}+\int\frac{d^{3}\mathbf{q}}{\left(  2\pi\right)  ^{3}}\left(
\frac{g_{\mathbf{q}}^{2}}{\mu-\epsilon_{\mathbf{q}}-\omega_{\mathbf{q}}}%
+\frac{m_{FB}U_{FB}^{2}}{2n_{B}^{-1}m_{F}\epsilon_{\mathbf{q}}}\right)  .
\label{chemical potential}%
\end{equation}
It has been well established that highly imbalanced Fermi systems support not
only attractive
\cite{prokofev08PhysRevB.77.020408,mora09PhysRevA.80.033607,schirotzek09PhysRevLett.102.230402,nascimbene09PhysRevLett.103.170402}
but also repulsive
\cite{cui10PhysRevA.81.041602,massignan11EurPhysJD.65.83,schmidt11PhysRevA.83.063620,marco12Nature.485.619, kohstall}
polarons. Likewise, repulsive polarons are also expected to exist in our
system when $U_{BF}$ is tuned on the repulsive side of the Feshbach resonance,
but the repulsive polarons are highly unstable \cite{rath13arXiv:1308.3457}.
\ In this work, consistent with the assumption that led to the self-energy in
Eq.\ (\ref{self-energy}), we limit our investigation to systems with attractive
Bose-Fermi interaction ($U_{BF}<0$) where polarons are always attractive.
\ The $\omega$-intercept in Fig.\ \ref{Fig3:spectralFunction}(a)\ indicates a
polaron energy (or rather a binding energy) of $E_{0}=-0.276$ $(\hbar
^{2}/m_{F}\xi_{B}^{2})$, which is in agreement with the one determined from
Eq.\ (\ref{chemical potential}). 

The existence of a negative polaron energy creates, for small momenta, a gap
between the polaron peak dispersion curve and the impurity-phonon scattering
continuum (light blue region), a region defined to be above $\omega_{c}\left(
\mathbf{p}\right)  =\min$ $\left(  \epsilon_{\mathbf{p}-\mathbf{q}}%
+\omega_{\mathbf{q}}\right)  $, where $\mathbf{q}$ covers all the possible
phonon momenta. For an impurity with a momentum below a certain threshold
$p_{c}$ [$\approx 0.88$ in Fig.\ \ref{Fig3:spectralFunction}(a)], the
impurity-phonon interaction is coherent, and the spectral function describing
the peak is then a $\delta$-like function as in the case of free fermions.
\ For an impurity with a momentum above $p_{c}$ the polaron peak line enters
the continuum where the phonon states become energetically available for
impurity scattering (and at very large momenta, as expected, the polaron peak
line asymptotes to the free impurity dispersion). The impurity scattering
involving the phonon emissions then causes the peak line to be broadened as
illustrated in diagrams on the right panel of Fig.\ \ref{Fig3:spectralFunction}.

It is well-known that the gap shown in Fig.\ \ref{Fig3:spectralFunction}(a) can
cause the calculation using Fermi liquid theory (equivalent to the
Tamm-Dancoff approximation) to be quite different from the calculation using
the on-mass-shell approximation (also known as the Rayleigh-Schr\"{o}dinger
approximation) \cite{mahan00ManyParticlePhysicsBook}.  In the
present problem, this same gap has an added consequence of diminishing the
role that anisotropy plays in Fermi liquid theory. In a future study, we
plan to tailor \red{Feynman's} path integral formalism
\cite{feynman55PhysRev.97.660} for exploring the polaron physics in
our model; as a superior all coupling approximation, Feynman's method would
not only extend the current study to the strong-coupling regime, but would
also help arbitrate between the two approaches just mentioned.

\section{Conclusion}

In conclusion, we have considered a polaronic model in which impurity fermions
interact, via the short-range s-wave scattering potential, with background
bosons in a dipolar condensate where bosons interact via both short-and
long-range interactions.\ We have described such a model with a Fr\"{o}hlich
type of Hamiltonian where the role of phonons is played by density
fluctuations of dipolar bosons. The polaron in this model emerges as an
impurity dressed with a cloud of phonons which, due to the competition between
the attractive and repulsive part of the dipole-dipole interaction, obey an
anisotropic dispersion spectrum. Taking into consideration only the
single-phonon-impurity scattering processes, we have constructed a self-energy
capable of capturing the main polaronic features of our model operating in the
weak coupling regime. We have described the polaron using the self-energy on
the on-mass-shell, discussing, in particular, how anisotropy affects the
\v{C}erenkov radiation of Bogoliubov phonon modes, which can be directly
verified by experiments in which a dipolar BEC moves against an obstacle. We
have also described the polaron in the spirit of Fermi liquid theory, focusing
on the spectral function for the impurity fermions, which is directly
accessible to the momentum resolved rf spectroscopy in cold atoms.
\ Finally, we stress that while this work concentrates on the
three-dimensional dipolar Bose-Fermi mixture, it can be generalized to its
two-dimensional analog where the polaron physics may be greatly enriched by
the existence of a \textquotedblleft roton\textquotedblright\ minimum
structure in the phonon spectrum of the dipolar BEC
\cite{santos03PhysRevLett.90.250403,ticknor11PhysRevLett.106.065301}.

\textit{Note added:} While this work was under review, we were informed of the article \cite{Shashi14arXiv:1401.0952} by its authors, which uses a variational approach to calculate the spectral properties and rf spectroscopy of polarons in a \textit{nondipolar} BEC.

\section{Acknowledgements}

H.\ Y.\ L.\ is supported in part by the US Army Research Office under Grant No.\
W911NF-10-1-0096 and in part by the US National Science Foundation under Grant
No.\ PHY11-25915.


\end{document}